\documentstyle[12pt]{article}
\begin{document}
\baselineskip = 21 pt
\thispagestyle{empty}
\title{
\vspace{-2.5cm}
\begin{flushright}
\begin{tabular}{c c}
 & { \normalsize MPI-Ph/92-67}\\
 & {\normalsize August 1992}
\end{tabular}
\end{flushright}
\vspace{1.5cm}
On the Finite Temperature Effective
Potential in Scalar QED with N Flavors.
\\
 ~\\}
\author{{\bf
 M. Carena}  and {\bf C. E. M. Wagner} \\
 ~\\
Max Planck Institut f\"{u}r Physik\\
Werner Heisenberg Institut\\
F\"{o}hringer Ring 6\\
D-8000 M\"{u}nchen 40, Fed. Rep. of Germany.\\
{}~\\
{}~\\
{}~\\
{}~\\
{}~\\ }
\date{
\begin{abstract}
The effective potential of scalar
quantum electrodynamics with N flavors of complex
scalar fields is studied, by performing a self consistent 1/N expansion up
to next to leading order in $1/N$. Starting from the broken phase at zero
temperature,  the theory exhibits a phase transition to the
symmetric
phase at some finite temperature $T_c$.
We work in general covariant gauges and
demonstrate the gauge invariance of both, the critical temperature
$T_c$ and the
minimization condition at any finite temperature T. Furthermore, the
only minimum
of the potential is at zero scalar vacuum expectation value for any
temperature $T \geq T_c$ and varies continuously to nonzero values
for temperatures below $T_c$, implying the existence of a second order
phase transition.
\end{abstract}}
\maketitle
\newpage
\section{Introduction}

In the last years, there has been a great
interest in understanding the nature of the finite temperature
symmetry restoration
phase transition  in the Standard Model (SM) \cite{KL} - \cite{improved5}.
Recently,
the main motivation arose
from the realization that
the rate for baryon number violation
in the SM at finite
temperature is much larger than what it
was originally assumed \cite{KRS} - \cite{AM}.
 In fact, once
the temperature is much higher than the height of the energy barrier
between the different topologically distinct vacua \cite{KM},
the exponential
suppression of the anomalous baryon number violation processes,
present
at low temperature,  disappears \cite{KRS}-\cite{AM}.
Thus, at sufficiently high temperatures, the anomalous baryon
number violation processes could wash out the primordial baryogenesis
generated, in some grand unified scenarios, at the very early stages of
the universe \cite{Oxford}.
In such cases, the generation of baryon number could
have occurred through non-equilibrium processes in
the transition from the
symmetric to the broken $SU(2)_L
\times U(1)_Y$  phase \cite{KRS}.
Quite generally, models of electroweak
baryogenesis \cite{models}
fulfill  Sakharov's  three
requirements
\cite{Sakharov}
for the
generation of baryon asymmetry. Besides the nonconservation of B
the underlying dynamics must involve nonequilibrium processes and
violate CP.
The violation of
CP is  present in the SM and its extensions,
although in the SM it is probably too weak
in order to generate the required baryon asymmetry \cite{models}.
To satisfy the nonequilibrium condition, these
models rely on the presence of a first order phase transition,
sufficiently strong
in order to imply the suppression of baryon number violation
processes in the broken phase.

The nature of the electroweak phase transition
is a fundamental
question which needs    to be carefully explored and requires
a detailed study of the effective potential of gauged scalar
theories at
finite temperature. A  one loop analysis of the abelian
theory was first done by Dolan and Jackiw
\cite{DJ}.
The one loop effective potential
shows the existence of a first order phase
transition, whose  strength  depends on the values of the
electroweak gauge
and the scalar quartic coupling constants \cite{KL}.
However, the naive one loop definition of the
effective potential is known to have serious infrared problems,
such as
the appearance of a complex effective potential even at
temperatures above the critical temperature, $T_c$, at which
the transition from the broken to the symmetric phase takes
place.
Moreover, it
is found that the critical temperature itself is a gauge dependent
quantity. It was realized in Ref.\cite{DJ}
 that both - the infrared
and the gauge dependence -
problems  do not appear, if  only the dominant
terms in the high temperature expansion
are kept. Within such an approximation,
a definition of a gauge invariant critical
temperature can be achieved by
assuming
the existence of a second order phase transition. However, in order
to study the nature of the  phase transition, higher order
terms in the effective potential  need to be computed.

 Quite recently, an improved  analysis of the
one loop effective potential
has been considered by several
authors \cite{improved} -\cite{improved5}.
The  improved one loop effective potential
is generically obtained by the inclusion of the
finite temperature one loop corrections, at zero external
momentum, to the
boson propagators. When this process is carefully
done, in order to avoid double counting of diagrams,
it is equivalent to the inclusion of the so called ring
diagrams in the one loop
computation \cite{improved3}.
Such resummation of diagrams solves several problems
inherent to
the one loop expansion
and leads to a first order phase transition, although
weaker than that one found within
the one loop expansion \cite{improved4}. However,
in the pure scalar case the ring expansion breaks down at temperatures
close to the critical one, $|T - T_c| \simeq
\tilde{\lambda} T_c$, with $\tilde{\lambda}$
the scalar quartic coupling \cite{improved5}.
Hence, it can not be used to analyze the behavior
of the effective potential at $T \simeq T_c$. In addition, after the inclusion
of the gauge fields, for values of the squared
gauge coupling $g^2$ of the order of the
quartic coupling it is not possible to conclude that the phase transition is
first order based solely on the ring expansion. This is due to the fact that,
within the ring approximation,
and for $g^2 \simeq \tilde{\lambda}$, the difference
between the transition temperature $T_0$, at which the curvature at the
origin vanishes,  and the critical temperature $T_c$
is of order $g^2 T_c$,
but the ring expansion
breaks down for $|T - T_c| \leq g^2 T_c$ \cite{improved5}.
Thus, an analysis beyond the improved
 one loop approximation is necessary in order to
study the nature of the phase
transition for $g^2/\tilde{\lambda} \simeq {\cal{O}}(1)$, which
is the phenomenologically interesting region in which
the Higgs mass is of the order of the gauge boson mass.

As it
has been explained in the seminal work by Dolan and Jackiw
 \cite{DJ}, for the
pure scalar case
a self consistent resummation of the diagrams contributing to the
effective potential can be done,
by extending the theory to include N flavors of self interacting
scalar fields
 and performing a 1/N study of this model. The 1/N expansion
provides information which does not
appear at any fixed order of perturbation
theory and, hence,
it is a more appropriate tool  to study
the nature of the phase transition, which is in
itself a nonperturbative phenomenon. Studying the behaviour of
the effective squared mass,
by doing a self consistent 1/N expansion, Dolan and Jackiw \cite{DJ}
showed that the infrared problems
inherent to
the pure O(N) scalar theory disappear.
The explicit form of the effective potential
at finite temperature was, however,
 not derived in this analysis, and it was  studied
in Refs. \cite{HW} and \cite{Vid}.
A similar resummation of diagrams has been recently performed
in Ref. \cite{CERN}, while an analysis of the $N$ component -
$\phi^4$ theory,  by a method
based on average fields, has been considered
in Ref.\cite{TW}.

The validity of the finite temperature computations
in the scalar O(N) linear model was questioned,
when it was realized that,
in the continuum limit,
the effective
scalar potential at zero temperature was either
complex, unstable or did not allow the spontaneous breakdown of the O(N)
symmetry \cite{CJP} - \cite{AKS}.
 This issue was clarified by Bardeen and Moshe Moshe
 \cite{BM} , who explained that
the problems appearing in the large N computations
were not associated with the 1/N expansion but with a more
fundamental nonperturbative property of the O(N) scalar model in four
dimensions, which is the issue of
triviality. In fact, for positive bare quartic coupling,
the continuum limit
of the four dimensional O(N) theory
is trivial,
in the sense that the renormalized coupling goes to zero when
the cutoff is removed.
This implies that,  to work with a nontrivial
theory, one has  either
to take a negative bare quartic coupling, rendering
the theory unstable
at finite temperature \cite{BM},
 or treat the theory as an effective one, by keeping
a finite effective cutoff. For a
large effective cutoff, the 1/N expansion at zero
and finite temperature
can be defined in a consistent way,
avoiding the problems mentioned above.
Thus, in the following, we shall assume that we are dealing with an
effective theory valid up to an  energy scale of the order of an
effective cutoff $\Lambda$,
at which, quite generally, new physics should appear.

It must be mentioned, that the
 nature of the phase transition for scalar QED with $N$
flavors and a scalar Higgs heavier than the gauge boson,
have been also studied by using a
$4-\epsilon$ expansion \cite{HLM}-\cite{Ginsparg}.
Within such approximation, the theory
may be analyzed
by performing a
renormalization group study for small $\epsilon$.
The
high temperature theory may be treated as  an effective
three dimensional theory and thus,
to obtain
physical results, the expansion parameter
must be taken to be $\epsilon = 1$. In this framework,
the critical value for the
number of flavors,
$N_c$, below which the phase transition becomes first
order, was found to be surprisingly large,
$ N_c = 183$. Based also on this analysis, Halperin, Lubensky
and Ma \cite{HLM} predicted a first order phase transition
for the smectic-A to nematic phase transition in liquid
crystals.
However, when contrasted with experiment, this
transition was
found to be second order \cite{DH}. In addition, for
$\epsilon \rightarrow 1$, a
($2 + \epsilon$) expansion
gives results which are at variance
with those ones coming from
the ($4 - \epsilon$)
expansion \cite{HLM},\cite{Russel}. In particular,
in the case in which the Higgs  particle is
heavier than the gauge boson, the $(2 + \epsilon)$ expansion
indicates a much lower critical value of
$N$, if any, for the three dimensional theory
\cite{Russel}.

This paper is organized as follows: In section 2
we present the model,
a scalar theory with N flavors of complex scalar fields coupled to
an abelian gauge field.
In section 3 we derive the
general
expression of the effective potential up to next to leading
order in 1/N.
In section 4  we
concentrate on
the leading order results,
and we show
that a second order phase transition takes
place in this case.
In section 5 we carry out the analysis of the critical
temperature and the nature of the
phase transition up to
 next to leading order in
1/N. We start
discussing  three related aspects of the theory,
which are the renormalization of couplings, the triviality of
the theory and the absence of tachyons. Then, we analyze
the radiative corrections
at finite temperature to finally perform the
minimization of the
effective potential up to
next to leading order in $1/N$.
We demonstrate that a gauge invariant expression
for the position of the minimum is obtained. In addition, the
position of the minimum is not modified by next to leading order
corrections and, hence, the phase transition remains second
order. A comparison of
our large $N$
results with those  previously derived
within the improved one loop method is performed in section 6,
where we analyze the source of the  discrepancy
between the results obtained in both approaches.
We reserve section 7 for our conclusions.

\section{ The  Model }

 The Lagrangian density for a scalar
theory with N flavors of complex scalar fields in interaction with
an abelian gauge field is given by
\begin{equation}
{\cal{L}} = -\frac{1}{4} F_{\mu \nu} F^{\mu \nu} - \frac{1}{2 \alpha}
\left( \partial_{\mu} A^{\mu} \right)^2  + \frac{1}{2}
\left( {\cal{D}}_{\mu} \phi^a \right)^{\dagger}
\left( {\cal{D}}^{\mu} \phi^a \right) - \frac{
\tilde{\lambda}}{4!} \left( |\phi|^2 - v^2 \right)^2 .
\end{equation}
In the above expression,
$a$ is a flavor index taking values from 1 to $N$, $\phi^a$ is
a complex scalar field, $|\phi|^2 = \phi_a^{\dagger} \phi_a$,
$A_{\mu}$ is an abelian gauge field,
and  ${\cal{D}}_{\mu}$  is the covariant derivative
associated with it. In addition,
we have included a  gauge fixing term
with $\alpha$ being the gauge fixing parameter
($\alpha = 0$ is the Landau gauge). Note that, at vanishing gauge coupling,
this model reduces to the O(2N) linear sigma model.

Following the method first proposed by Jackiw \cite{Jackiw},
to compute
the effective potential $V(\hat{\phi})$ for this theory,
we shift the fields $\phi_a \rightarrow \hat{\phi}_a
+ \phi_a$, where
$\hat{\phi}_a$ is the vacuum expectation value of the scalar field and
$\phi_a$ stands for its quantum fluctuations. The Lagrangian must be
truncated by dropping all terms which are linear in the $\phi_a$
quantum fluctuations. The new Lagrangian, ${\cal{L}}
(  \hat{\phi_a}
+ \phi_a, A_{\mu})$, may be decomposed into a term quadratic in
the $A_{\mu}$ and $\phi_a$ fields and
an interaction term. The effective
potential is given by the sum of the tree level potential plus the
one loop contributions, obtained from the bilinear part of the
shifted Lagrangian, plus higher order loop contributions,
given by
the sum over all connected
one-particle irreducible vacuum graphs for the theory
described by the shifted Lagrangian density ${\cal{L}}
(  \hat{\phi_a}
+ \phi_a, A_{\mu})$.

The quartic interactions pose a disadvantage for the diagrammatical
 analysis
of the theory. Therefore, following Refs. \cite{CJP} - \cite{Root},
we shall eliminate the scalar
quartic term through the introduction of an auxiliary field $\chi$.
The
modified Lagrangian reads
 \begin{equation}
{\cal{L}}(\phi_a,
A_{\mu},\chi) = {\cal{L}}(\phi_a,A_{\mu}) +
\frac{3}{2 \tilde{\lambda}} \left[ \chi  - \frac{\tilde{\lambda}}
{6} \left(
|\phi|^2 - v^2 \right) \right]^2.
\end{equation}
The Euler-Lagrange equation for $\chi$ is a
constraint equation relating $\chi$ with the scalar field $\phi$.
In fact, the effective
potential for the modified theory $V(\hat{\chi}, \hat{\phi_a})$
reduces  to the one of the original theory if
$\hat{\chi}$ satisfies the requirement
\begin{equation}
\frac{\partial V}{ \partial \hat \chi} = 0,
\end{equation}
which defines the vacuum expectation value
$\hat{\chi}$ as a function of $\hat{\phi}$.

A self consistent
$1/N$ expansion can be defined if
  the gauge and quartic coupling constants
$g$ and $\tilde{\lambda}$, respectively,
 depend on $N$ so that in the limit of
$N \rightarrow \infty$,
$g^2 N$ and $\tilde{\lambda} N$ go to constant values.
Hence, for the purpose
of our study it is better to define the new coupling constants
$e \equiv
g \sqrt{N}$ and $\lambda \equiv \tilde{\lambda} N$.
Once the field $\chi$ is introduced the Lagrangian density of the
theory reads
\begin{eqnarray}
{\cal{L}}(\phi_a, A_{\mu},\chi)
& = & -\frac{1}{4} F_{\mu \nu} F^{\mu \nu} - \frac{1}{2 \alpha}
\left( \partial_{\mu} A^{\mu} \right)^2  + \frac{1}{2}
\partial_{\mu} \phi_{a,i} \partial^{\mu} \phi_{a,i} -
\frac{e}{\sqrt{N}}
 A^{\mu} \epsilon^{i j} \phi_{a,j}
\partial_{\mu} \phi_{a,i}
\nonumber\\
& + & \frac{e^2}{2 N} A_{\mu} A^{\mu} \phi_{a,i} \phi_{a,i} +
\frac{3 N}{2 \lambda} \chi^2  - \frac{\chi}
{2} \left(
|\phi|^2 - v^2 \right),
\end{eqnarray}
where $i = {1,2}$, $\phi_{a,1}$ and
$\phi_{a,2}$  are the real and imaginary parts
of the field $\phi_a$, respectively,
and summation over $a$ and $i$
is understood. Although the introduction of the field $\chi$ does not
alter the dynamics of the full theory, it does lead to a new
perturbation series, in which the 1/N expansion has a simpler
diagrammatic interpretation.
After considering the shifted fields, which now
include the shift for the auxiliary field $\chi$,
\begin{equation}
\phi_{a} \rightarrow \sqrt{N} \hat{\phi_a} + \phi_a,
\;\;\;\;\;\;\;\;\;\;\;
\chi \rightarrow \hat{\chi} + \chi
\end{equation}
and ignoring the  linear terms in
the quantum fluctuations $\chi$ and $\phi_a$, the shifted
Lagrangian reads
\begin{eqnarray}
{\cal{L}}& = & - \frac{1}{4} F_{\mu \nu} F^{\mu\nu}
-\frac{1}{2 \alpha} \left( \partial_{\mu} A^{\mu} \right)^2
+ \frac{e^2}{2} A^{\mu} A_{\mu} \hat{\phi}_{a,i} \hat{\phi}_{a,i}
-e A^{\mu} \epsilon^{ij} \hat{\phi}_{a,j} \partial_{\mu} \phi_{a,i}
\nonumber\\
& - & \frac{e}{\sqrt{N}} A^{\mu} \epsilon^{ij} \phi_{a,j}
\partial_{\mu} \phi_{a,i} + \frac{e^2}{2N} A^{\mu} A_{\mu}
\phi_{a,i} \phi_{a,i} + \frac{e^2}{\sqrt{N}} A^{\mu} A_{\mu}
\hat{\phi}_{a,i} \phi_{a,i}
+ \frac{1}{2} \partial_{\mu} \phi_{a,i} \partial^{\mu} \phi_{a,i}
\nonumber\\
&+& \frac{3N}{2 \lambda} \chi^2 - \frac{1}{2} \hat{\chi} \phi_{a,i}
\phi_{a,i}
- \sqrt{N} \hat{\phi}_{a,i} \chi \phi_{a,i} -
\frac{1}{2} \phi_{a,i} \phi_{a,i} \chi  - V_{tree} ,
\label{eq:Lagr}
\end{eqnarray}
where the tree level potential is given by,
\begin{equation}
V_{tree} = - \frac{3 N}{2 \lambda} \hat{\chi}^2 +
\frac{N}{2} \hat{\chi} \left( |\hat{\phi} |^2 - v^2 \right) .
\end{equation}
In the above, we have followed Root \cite{Root} in
rescaling the vacuum expectation value
$\hat{\phi_a}$ and, consistently,
$v$ by a factor $\sqrt{N}$. This simplifies
the $1/N$ counting. In fact, if this rescaling were
not done, the  vacuum expectation value would be of order
N in the natural scales of the theory,
$\hat{\phi}^2 = N  F(\mu^2,T^2)$,
where $F$ is a function of the
bare mass $\mu^2 = - \tilde{\lambda} v^2/6$,
the temperature $T$ and the couplings of the theory.
This extra factor of
$N$ would have to be properly
considered while doing the expansion at
next to leading order. The physical results, of course,
would remain unchanged.

The above Lagrangian may be decomposed into a quadratic and an
interacting part.  The expression of the quadratic Lagrangian reads
\begin{eqnarray}
{\cal{L}}_{quad}& = & \frac{1}{2} A^{\mu} \left[ (\partial^2 +
e^2 \hat{\phi}^2 ) g_{\mu \nu}
- \partial_{\mu} \partial_{\nu} \left( 1 - \frac{1}{\alpha} \right)
\right] A^{\nu} - \frac{1}{2} \phi_{a,i} \left( \partial^2 +
\hat{\chi} \right) \phi_{a,i}
\nonumber\\
& + & \frac{3N}{2\lambda} \chi^2
- \sqrt{N} \hat{\phi}_{a,i} \chi \phi_{a,i}
-e A^{\mu} \epsilon^{ij}
\hat{\phi}_{a,j} \partial_{\mu} \phi_{a,i} \,
\end{eqnarray}
from which  it is easy to deduce the inverse propagator matrices
relevant for the computation of the effective potential.
The interacting part of the Lagrangian ${\cal{L}}_I$, on the
other hand,
determines the vertices of the boson fields interactions.
For simplicity,
we will assume in the following that
\begin{equation}
\hat{\phi}_{1,1} = \hat{\phi},
\end{equation}
while all other vacuum expectation values are zero. In fact,
it is not important which
state is chosen as the ground state of the theory.
Independent of such a choice the original global U(N) symmetry
is spontaneously broken to U(N-1) and 2N-1 Goldstone bosons
are generated, one of which is eaten by the gauge field
through the usual Higgs
mechanism.

We obtain two decoupled  inverse propagator matrices. One
that mixes the gauge field $A_{\mu}$ with $\phi_{1,2}$,
${\cal{M}}(A_{\mu},\phi_{1,2})$, and the other
one which mixes the auxiliary field $\chi$ with $\phi_{1,1}$,
${\cal{M}}(\chi,\phi_{1,1})$. Thus, in momentum space
representation,
the nonzero components of the inverse propagator matrices read,

\begin{eqnarray}
i D^{-1}_{\mu \nu}& = & - \left[ g_{\mu\nu}\left( k^2 - e^2 \hat{\phi}^2
\right) - k_{\mu} k_{\nu} ( 1 - \frac{1}{\alpha} ) \right]
\nonumber\\
i D^{-1}_{a,i;b,j}& = & \delta_{ab} \delta_{ij} (k^2 - \hat{\chi})
\nonumber\\
iD^{-1}_{\chi \chi}& = & \frac{3N}{\lambda}
\nonumber\\
iD^{-1}_{1,1;\chi} & = & - \hat{\phi} \sqrt{N}
\nonumber\\
iD^{-1}_{1,2;\mu} & = & i e k_{\mu} \hat{\phi} .
\label{eq:invprop}
\end{eqnarray}
Considering the above expressions it is straightforward to compute the
determinant of the inverse propagator mass matrices,
which in Euclidean
space are given by
\begin{eqnarray}
\det {\cal{M}}(A_{\mu}, \phi_{1,2}) = \frac{
(k^2 + \hat{\chi}) (k^2 + e^2 \hat{\phi}^2 )^3}{ \alpha}
\left[
k^2 + \frac{e^2 \hat{\phi}^2 \hat{\chi} \alpha}{(k^2 + \hat{\chi})}
\right],
\label{eq:det1}
\end{eqnarray}
\begin{eqnarray}
\det {\cal{M}}(\chi, \phi_{1,1}) = \frac{-3N}{\lambda}
\left[k^2 + \hat{\chi} + \frac{\lambda}{3} \hat{\phi}^2 \right].
\label{eq:det2}
\end{eqnarray}
Furthermore,
the propagators of the scalars and gauge bosons, which are relevant
for the computation of the effective potential up to next to
leading order in $1/N$ may be then obtained from
Eqs. (\ref{eq:invprop}),
(\ref{eq:det1}) and (\ref{eq:det2}), and in Euclidean space
are given by
\begin{equation}
iD_{\mu\nu}  =  \left( \delta_{\mu\nu}
- \frac{k_{\mu} k_{\nu}}{k^2}
\right) \frac{1}{k^2 + e^2 \hat{\phi}^2}
+ \frac{k_{\mu} k_{\nu}}{k^2} \frac{ \alpha \left(k^2 + \hat{\chi}
\right)}{\left[ k^4 + \hat{\chi} k^2 + \alpha \hat{\chi} e^2
\hat{\phi}^2
\right]}
\label{eq:daa}
\end{equation}
\begin{equation}
i D_{\chi\chi} = - \frac{ \left(k^2 + \hat{\chi} \right)
\lambda/3N}{ \left[ k^2 + \hat{\chi} +
\lambda \hat{\phi}^2/3
\right] }
\label{eq:dchichi}
\end{equation}

\section{The Effective Potential}

As we mentioned above,
in order to analyze which Feynman diagrams contribute to the
effective potential up to next to leading order in $1/N$, it is
necessary to consider the behaviour in $1/N$ of the propagators
and of  the interaction vertices. From
Eqs. (\ref{eq:invprop})-(\ref{eq:dchichi})
we observe that a factor
$1/N$ is associated with $D_{\chi\chi}$ and a factor $1/\sqrt{N}$
is associated with $D_{1,1;\chi}$. Moreover, a factor of order
$N$ is associated with any closed loop of $\phi_a$
fields, due
to the summation over all possible internal fields. From the
interaction part of the Lagrangian in  Eq.(\ref{eq:Lagr}),
we observe that a factor
$1/\sqrt{N}$ is associated with the derivative coupling of the
scalar fields $\phi_a$ to the gauge bosons,
 while a factor $1/N$
appears in any quartic vertex $A^2 \phi_{a,i}
 \phi_{a,i}$. Note that the vertex $\chi \phi_{a,i} \phi_{a,i}$
has, instead, a factor 1 in the coupling.

The only diagram contributing to the potential at
leading order in $1/N$ \cite{HW},\cite{Vid},
$V_{l.o.}$, is the closed loop involving the
2(N-1) Goldstone bosons which do not mix with either
$A_{\mu}$ or $\chi$,
\begin{equation}
V_{l.o.} =  \frac{2 (N-1)}{2}
\int_k
\ln \left( k^2
+ \hat{\chi} \right).
\end{equation}
We perform the finite temperature computations in the
imaginary time formalism: After a Wick rotation to Euclidean
space we impose periodic boundary conditions on the
time direction of length $L = T^{-1} \equiv \beta$. For
simplicity of notation, we have defined
\begin{equation}
\int_k f(k) \equiv \frac{1}{\beta} \sum_{n=-\infty}^{n=\infty}
\int \frac{d^3 k}{(2\pi)^3} f(\omega_n, \vec{k}),
\end{equation}
with $\omega_n = 2 \pi n/ \beta$.

The computation of the effective potential at next
to leading order in $1/N$ involves two type of contributions.
There are one loop contributions,
involving the two nontrivial propagator matrices,
\begin{equation}
V_{n.l.o}^{1.l.} =\frac{1}{2}
 \int_k \left[
\ln \det\left(
{\cal{M}}(A_{\mu}, \phi_{1,2})\right)
+ \ln \det\left( {\cal{M}}(\chi,\phi_{1,1})\right) \right] ,
\end{equation}
and there are
those coming from multiloop one particle irreducible graphs,
depicted in Fig.1. Observe that any
multiloop contribution
involving the mixing of $\chi$ and $A_{\mu}$, like those
depicted in Fig.2, vanish in the regularized theory.
After some work,
 the resummation of all the multiloop diagrams
contributing to next to leading order in $1/N$ gives
\begin{eqnarray}
V_{n.l.o}^{n-loop} & = & - \sum_{n=1}^{\infty} \frac{1}{2n}
\int_k \left[ - \frac{ \lambda \left(k^2 + \hat{\chi} \right)}
{3 \left[ k^2 + \hat{\chi} + \lambda \hat{\phi}^2/3 \right] }
\right]^n B^n(k^2,\hat{\chi})
\nonumber\\
& - & \sum_{n=1}^{\infty} \frac{1}{2n}
\int_k Tr \left[ i e^2
D_{\mu \nu} \Pi^{\nu \alpha} \right]^n ,
\end{eqnarray}
which can be rewritten as
\begin{eqnarray}
V_{n.l.o}^{n-loop} & = & \frac{1}{2} \int_k
\ln \left[ 1 + \frac{ \lambda \left( k^2 + \hat{\chi} \right)}
{3 \left( k^2 + \hat{\chi} + \lambda
\hat{\phi}^2 /3 \right)}
B(k^2,\hat{\chi}) \right]
\nonumber\\
& + & \frac{1}{2} \int_k Tr \ln \left( \delta_{\mu}^{\alpha}
- i e^2 D_{\mu \nu} \Pi^{\nu \alpha} \right),
\end{eqnarray}
where
\begin{equation}
B(k^2, \hat{\chi} ) = \int_p  \frac{1}{ \left( p^2 + \hat{\chi}
\right) \left[ \left( k + p \right)^2 + \hat{\chi} \right]}
\end{equation}
and
\begin{equation}
\Pi^{\nu \alpha}(k, \hat{\chi})
= \int_p \frac{ \left( 2 p + k \right)^{\nu}
\left( 2 p + k \right)^{\alpha} }{ \left( p^2 + \hat{\chi}
\right) \left[ \left( p + k \right)^2 + \hat{\chi} \right]}
- 2 \delta^{\nu \alpha} \int_p \frac{1}{p^2 + \hat{\chi}},
\end{equation}
is the finite temperature vacuum polarization contribution.
Furthermore, using the properties
\begin{equation}
\det {\cal{M}}(\phi_{1,2},A_{\mu}) = \left( k^2 + \hat{\chi} \right)
\det\left[ -i D^{-1}_{\mu \nu} \right]
\end{equation}
and
\begin{equation}
\det {\cal{M}}(\phi_{1,1}, \chi ) = \left( k^2 + \hat{\chi} \right)
D_{\chi \chi}^{-1},
\end{equation}
we arrive to the final formal expression for the effective potential
up to  next to leading order in $1/N$,
\begin{eqnarray}
V(\hat{\phi},\hat{\chi}) & = & V_{tree} + V_{l.o.} +
V_{n.l.o.}^{1.l.} + V_{n.l.o.}^{n-loop}
\nonumber\\
& = & - \frac{3N}{2 \lambda} \hat{\chi}^2 + \frac{N}{2}
\hat{\chi} \left( | \hat{\phi} |^2 - v^2 \right) +
\frac{ (2N) }{2} \int_k \ln \left( k^2 + \hat{\chi} \right)
\nonumber\\
& + & \frac{1}{2} \int_k \ln \det \left[
-i D^{-1}_{\mu \nu}(k,\hat{\chi}, \hat{\phi})
-
e^2 \Pi_{\mu \nu}(k,\hat{\chi}) \right]
\nonumber\\
& + & \frac{1}{2} \int_k \ln \left[
\frac{ \left(k^2 + \hat{\chi} \right) \left( 1 + \lambda
B(\hat{\chi}, k^2)/3
\right) + \lambda \hat{\phi}^2/3}{k^2 + \hat{\chi}} \right].
\label{eq:effpot}
\end{eqnarray}

As we discussed above, the expression for the effective potential
of the original theory
may be obtained from the above equation by imposing the
condition
\begin{equation}
\frac{\partial V (\hat{\phi},\hat{\chi})}{\partial \hat{\chi}} = 0,
\label{eq:gapeq}
\end{equation}
which determines the
expression of $\hat{\chi}$ as a function of $\hat{\phi}^2$.
Furthermore, as it was
first noticed in Ref. \cite{Root}, in order
 to obtain the effective
potential up to next to leading order in $1/N$,
 it is sufficient to
solve the gap equation,
Eq. (\ref{eq:gapeq}), up to leading order in $1/N$.
In fact, calling $\hat{\chi}(\hat{\phi}^2)$
the exact solution to the gap equation
and $\bar{\chi}(\hat{\phi}^2)$ the leading order solution,
then,
\begin{equation}
\hat{\chi}(\hat{\phi}^2) -
\bar{\chi}(\hat{\phi}^2) = {\cal{O}}(1/N).
\end{equation}
Expanding the effective potential around
$\hat{\chi}(\hat{\phi}^2) =
\bar{\chi}(\hat{\phi}^2)$, we have
\begin{equation}
V(\hat{\phi},\hat{\chi}) = V(\hat{\phi}, \bar{\chi})
+ \frac{\partial V(\hat{\phi},\bar{\chi})}{\partial \bar{\chi}}
\left( \hat{\chi} - \bar{\chi} \right) + {\cal{O}}(1/N) .
\end{equation}
However,
since $\bar{\chi}(\hat{\phi})$ is the solution to the
gap equation
at leading order, it follows that
$\partial V_{l.o.}(\hat{\phi},
\bar{\chi}) / \partial \bar{\chi} = 0$. Therefore, the second term
in the above equation is also of order 1/N,
\begin{equation}
V(\hat{\phi},\hat{\chi}(\hat{\phi})) =
V(\hat{\phi},\bar{\chi}(\hat{\phi})) + {\cal{O}}(1/N).
\end{equation}
Thus, in order to compute the effective potential up to next
to leading order in $1/N$ we just
need to consider
Eq.(\ref{eq:effpot}) with
$\hat{\chi}$ given by its leading order expression,
$\hat{\chi}(\hat{\phi}) = \bar{\chi}(\hat{\phi})$.

Moreover, for the extent of this work we shall concentrate in
computing the location of the extrema of the effective potential,
which are derived  from the condition,
\begin{equation}
\frac{ d V(\hat{\phi},\bar{\chi})}{d \hat{\phi}^2}
= 0.
\label{eq:minimum}
\end{equation}
In fact, the number and location of the extrema of
the effective potential
provide
sufficient information to study the nature of the phase
transition. In addition, as we will show below, a gauge
independent expression for the solutions to
Eq.(\ref{eq:minimum})
is found. Observe that
if   Eq.(\ref{eq:minimum}) is
not fulfilled for any nontrivial value of the scalar field,
the only minimum would be at the
origin and the global $U(N)$ symmetry,
together with the local $U(1)$ symmetry of the theory would be
preserved. From Eq.(\ref{eq:effpot}), we find the relation
$\partial{ V_{l.o.} } / { \partial\hat{\phi}^2 } = N \bar{\chi}/2$,
and recalling the fact that $\partial V_{l.o.}/
\partial \bar{\chi} = 0$, we obtain
\begin{eqnarray}
\frac{ dV }{ d \hat{\phi}^2} & = &
\frac{ N \bar{\chi}(\hat{\phi}^2)}{2}
\nonumber\\
& + & \left( \frac{\partial}{\partial \hat{\phi}^2 }
+ \frac{ \partial \bar{\chi}}{\partial \hat{\phi}^2}
\frac{\partial}{\partial \bar{\chi}}\right)
\left\{
\frac{1}{2} \int_k \ln \det \left[ -i D^{-1}_{\mu \nu}(k,\bar{\chi},
\hat{\phi}) -
e^2 \Pi_{\mu \nu}(k,\bar{\chi}) \right] \right.
\nonumber\\
& + & \left. \frac{1}{2} \int_k \ln \left[
\frac{ \left(k^2 + \bar{\chi} \right) \left( 1 + \lambda
B(\bar{\chi}, k^2)/3
\right) + \lambda \hat{\phi}^2/3}{k^2 + \bar{\chi}} \right]
\right\},
\label{eq:derpot}
\end{eqnarray}
which is the formal expression from which, after proper
integration and renormalization procedures, we shall be able to
determine
the critical temperature and the order of the phase transition
up to next to leading order in $1/N$.

\section{ Critical Temperature and Nature of the Phase Transition
at Leading Order in $1/N$}

In  section 3  we derived the expression of the effective
potential up to leading order in $1/N$ to be,
\begin{equation}
V(\bar{\chi},\hat{\phi}) = - \frac{ 3N}{2 \lambda_0}
\bar{\chi}^2 + \frac{ N }{2} \bar{\chi} \left( \hat{\phi}^2
- v_0^2 \right) + N \int_k \ln \left( k^2 + \bar{\chi} \right),
\end{equation}
where we have introduced
the subscript zero to denote
unrenormalized quantities. Thus, at leading order in 1/N, the
gap equation, which determines the expression of
$\bar{\chi}$ as a function of $\hat{\phi}^2$, reads
\begin{equation}
\frac{ \partial V }{ \partial \bar{ \chi }} =
- \frac{3N}{\lambda_0} \bar{\chi} + \frac{N}{2}
\left( \hat{\phi}^2 - v_0^2 \right) + N \int_k
\frac{1}{k^2 + \bar{\chi}} = 0.
\label{eq:leadgap}
\end{equation}
In order to explicitly evaluate the temperature dependent
integrals, we shall use the relation \cite{GPY}
\begin{equation}
\int_k f(k_0,\vec{k}) = \int \frac{d^4 k}{(2\pi)^4} f(k)
+\int_{-\infty + i \epsilon}^{\infty + i \epsilon}
\frac{dk_0}{2\pi} \int \frac{d^3k}{(2\pi)^3}
\left[ \frac{ f(k_0,\vec{k}) + f(-k_0,\vec{k})}{
\left( \exp(-i\beta k_0) - 1 \right)} \right],
\label{eq:FTtrick}
\end{equation}
which allows a decomposition in the zero temperature and
the finite temperature contributions - the first and second
terms in the right hand side
of the above  equation, respectively.
For the particular case
of the integral appearing in Eq.(\ref{eq:leadgap}), we
obtain,
\begin{equation}
\int_k \frac{1}{k^2 + \bar{\chi}} =
\int \frac{d^4 k}{(2 \pi)^4} \frac{1}{(k^2 + \bar{\chi})}
+ \int \frac{d^3 k}{(2 \pi)^3}  \frac{1}{ \sqrt{ \vec{k}^2 +
\bar{\chi}}
\left[\exp
\left(
\beta \sqrt{ \vec{k}^2 + \bar{\chi} } \right)
-1
\right]} ,
\end{equation}
where the $k_0$ dependence in the
 last integral was evaluated by performing a contour
integration in the complex $k_0$ plane. At large temperatures,
$T \gg \bar{\chi}$, the above integral may be expanded in a
way first derived by Dolan and Jackiw  \cite{DJ}
\begin{equation}
\int \frac{d^3 k}{(2 \pi)^3}  \frac{1}{ \sqrt{ \vec{k}^2 +
\bar{\chi}}
\left[
\exp \left( \beta \sqrt{ \vec{k}^2 + \bar{\chi} } \right)
-1 \right]} = \frac{1}{12 \beta^2} - \frac{\sqrt{\bar{\chi}}}
{4 \pi \beta} + \frac{\bar{\chi}}{16 \pi^2} \ln \left(
\frac{ c T^2}{\bar{\chi}} \right) + h.o.(\beta^2 \bar{\chi})
\label{eq:DJ}
\end{equation}
where $c = 16 \pi^2 \exp(1 - 2 \gamma)$, with $\gamma \approx 0.577$.

Moreover,
the zero temperature contribution is quadratically divergent and
needs to be regularized. Introducing a momentum cutoff $\Lambda$,
so that
\begin{equation}
\int \frac{d^4 k}{(2 \pi)^4} \frac{1}{ \left( k^2 + \bar{\chi}
\right) } = \frac{1}{16 \pi^2} \left[ \Lambda^2 - \bar{\chi}
\ln \left( \frac{ \Lambda^2 }{ \bar{\chi} }\right) \right],
\end{equation}
we then absorb the quadratic and logarithmic divergences by
defining the renormalized quantities

\begin{equation}
v^2 = v_0^2 - \frac{ \Lambda^2 }{8 \pi^2}
\end{equation}
and
\begin{equation}
\frac{1}{\lambda} = \frac{1}{\lambda_0}
+ \frac{1}{48 \pi^2} \ln \left[ \frac{\Lambda^2}{M^2} \right],
\label{eq:renlam}
\end{equation}
where $M^2$ is a renormalization scale. Observe that, for any
positive value of the unrenormalized quartic coupling $\lambda_0$,
the renormalized coupling $\lambda$ vanishes at
$\Lambda \rightarrow \infty$, and the theory
becomes trivial
in the continuum limit.
If the unrenormalized quartic coupling
is taken to be negative, the theory becomes unstable
\cite{BM}. Therefore, the only consistent definition
 of the theory is
when it is considered as an effective
theory with an effective finite
cutoff $\Lambda$.

 We can further define a
temperature dependent coupling by
\begin{equation}
 \frac{1}{\lambda_T} =
\frac{1}{\lambda} - \frac{1}{48 \pi^2} \ln \left[\frac{c T^2}
{M^2} \right]
\end{equation}
to finally rewrite the gap equation as
\begin{equation}
\bar{\chi} - \frac{\lambda_T}{6} \left( \hat{\phi}^2
- v^2 + \frac{1}{6 \beta^2} \right) +
\frac{\lambda_T}{12 \pi \beta} \sqrt{\bar{\chi}} = 0 .
\label{eq:leadgap2}
\end{equation}
The above expression gives a quadratic equation in
$\sqrt{\bar{\chi}}$, which can be
easily solved to find \cite{DJ}, \cite{Vid}
\begin{equation}
\sqrt{\bar{\chi}} = - \frac{\lambda_T}{ 24 \pi \beta}
+ \sqrt{ \left( \frac{ \lambda_T }{ 24 \pi \beta }
\right)^2 + \frac{ \lambda_T}{6} \left( \hat{\phi}^2
- v^2 + \frac{1}{6 \beta^2} \right) },
\label{eq:sqrtchi}
\end{equation}
where the plus sign has been chosen in order to obtain positive
values of $\sqrt{\bar{\chi}}$, which is a precondition for the
validity of the gap equation derived above.

As we have shown in section 3, the
minimization condition up to leading order in $1/N$ reads,
\begin{equation}
2 \frac{dV(\hat{\phi})}{d \hat{\phi}^2}  \equiv N \bar{\chi}(\hat{
\phi}^2 ) = 0.
\end{equation}
Considering the explicit expression for $\bar{\chi}(\hat{\phi}^2)$,
given in Eq.(\ref{eq:sqrtchi}), this yields the relation
\begin{equation}
\hat{\phi}^2 - v^2 + \frac{1}{6 \beta^2} = 0,
\label{eq:phimini}
\end{equation}
which determines the value of $\hat{\phi}$ at the minimum.
Eq. (\ref{eq:phimini})
has no solution for temperatures above the critical
value
\begin{equation}
T_c^2 \equiv
\frac{1}{\beta_c^2} = 6 v^2 ,
\end{equation}
and, hence, the minimum is at $\hat{\phi}_{min}
= 0$  in such region of $T$.\footnote{
Observe that there is a difference in a
factor 2  with respect to  the results
of Refs. \cite{DJ},\cite{HW},\cite{Vid}
due to the presence of $2N$, instead of $N$, real scalar
bosons in the theory.}
On the other hand,
at temperatures below the critical one,
the vacuum expectation
value is given by
\begin{equation}
\hat{\phi}_{min}^2 = \frac{1}{6} \left( \frac{1}{\beta_c^2} -
\frac{1}{\beta^2} \right).
\label{eq:minevo}
\end{equation}
Equation (\ref{eq:minevo})
shows that the vacuum expectation value varies
continuously from zero to nonzero values, signaling the
presence of a second order phase transition within the
leading order in $1/N$ solution of the model.
It is worth to remark that, for values of
$\hat{\phi}^2$ at the left of
the minimum, $\hat{\phi}^2 < \left(v^2 - 1/6\beta^2\right)$,
$\sqrt{\bar{\chi}}$ takes negative or even complex values,
showing that, as was first discussed by Coleman, Jackiw and
Politzer \cite{CJP},
the effective potential can not be defined in
a sensible way in that region.

\section{Critical Temperature and Phase Transition: Analysis
up to Next to Leading Order in 1/N.}

In terms of the renormalized quantities defined in
section 4, and within the framework of the high temperature
expansion,
we can rewrite the effective potential up to
next to leading order in $1/N$, Eq.(\ref{eq:effpot}), as
\begin{eqnarray}
V(\hat{\phi},\bar{\chi}) & = & - \frac{3 N}{2 \lambda_T}
\bar{\chi}^2 + \frac{N}{2} \bar{\chi} \left( \hat{\phi}^2
- v^2 \right) + \frac{ N \bar{\chi} }{12 \beta^2} -
\frac{ N \bar{\chi}^{3/2} }{6 \pi \beta}
\nonumber\\
& + & \frac{1}{2} \int_k \ln \det \left[
-i D^{-1}_{\mu \nu}(k,\bar{\chi},\hat{\phi})
- e_0^2 \Pi_{\mu \nu}(k,\bar{\chi}) \right]
\nonumber\\
& + &
\frac{1}{2} \int_k \ln \left[ \frac{ \left( k^2 + \bar{\chi}
\right)\left( 1  + \lambda_0 B(\bar{\chi},k)/3 \right)
+ \lambda_0 \hat{\phi}^2/3 }{ k^2 + \bar{\chi} } \right].
\label{eq:nextpot}
\end{eqnarray}
As we shall explicitly show in
section 5.1,
$B(\bar{\chi},k^2)$ and $\Pi_{\mu \nu}
\left(\bar{\chi}, k^2 \right)$ are logarithmically divergent
quantities. However, these logarithmic divergences can be
naturally absorbed in the renormalization of the gauge and
quartic couplings entering in the integrands of
Eq.(\ref{eq:nextpot}). The
remaining divergences, arising from the zero temperature
contributions
to the integrals, may be absorbed in the
renormalization of  $\lambda$ and $v^2$ at
next to leading order in $1/N$.

{}From Eq.(\ref{eq:derpot}) the minimization condition is given by
\begin{eqnarray}
\bar{\chi} & = & -\frac{1}{N} \left\{ \left[ \frac{\partial}{
\partial \hat{\phi}^2 } + \frac{\partial \bar{\chi}}
{\partial \hat{\phi}^2 }\frac{ \partial}{\partial \bar{\chi}}
\right] \right.
\nonumber\\
& &
\left.
\left(  \int_k \ln \det \left[
-i D^{-1}_{\mu \nu}(k,\bar{\chi},\hat{\phi})
- e_0^2 \Pi_{\mu \nu}(k,\bar{\chi}) \right] \right. \right.
\nonumber\\
& + & \left. \left.
\int_k \ln \left[ \frac{ \left( k^2 + \bar{\chi}
\right)\left( 1  + \lambda_0 B(\bar{\chi},k)/3 \right)
+ \lambda_0 \hat{\phi}^2/3 }{ k^2 + \bar{\chi} } \right]
\right) \right\}_{\bar{\chi} = 0}.
\label{eq:min}
\end{eqnarray}
Thus, since the minimum is obtained for $\bar{\chi} \simeq
{\cal{O}}(1/N)$, we can safely set $\bar{\chi} = 0$ in the right
hand side of Eq.(\ref{eq:min}),
under the assumption that no infrared
divergences arise in this process.

\subsection{Tachyon Poles and Triviality of the Gauged $O(2N)$ Model
at Zero $T$}

At zero temperature, it is possible
to compute the radiative corrections $B(p,\bar{\chi})$
and $\Pi_{\mu \nu}(p,\bar{\chi})$ by making use of a gauge invariant
regularization scheme, like Pauli - Villars. This gives the result
\begin{eqnarray}
B_{T=0}(p^2,\bar{\chi}(\hat{\phi})) & = & \frac{1}{16 \pi^2}
\left\{ \ln \left( \frac{\Lambda^2}{\bar{\chi}}\right)
\right.
\nonumber\\
& - &
\left.
2 \left[ \left( \frac{ 4 \bar{\chi} + p^2 }{ 4 p^2} \right)^{1/2}
\ln \left[ \frac{p^2}{\bar{\chi}} \left\{ \left(
\frac{ 4 \bar{\chi} + p^2 }{ 4 p^2} \right)^{1/2} + \frac{1}{2}
\right\}^2 \right] - 1 \right] \right\}
\end{eqnarray}
while
\begin{eqnarray}
\Pi^{\mu\nu}_{T=0} (p^2, \bar{\chi}(\hat{\phi}^2))
& = & \frac{ \left( \delta^{\mu \nu} p^2 - p^{\mu} p^{\nu}
\right)}{48 \pi^2} \left\{ -\ln \frac{\Lambda^2}{\bar{\chi}}
-\frac{2}{3} \right.
\label{eq:pimunut0}
\\
& + &  \left.
2 \left( \frac{ 4 \bar{\chi} + p^2 }{ p^2} \right)
\left[ \left( \frac{ 4 \bar{\chi} + p^2 }{ 4 p^2} \right)^{1/2}
\ln \left[ \frac{p^2}{\bar{\chi}} \left\{ \left(
\frac{ 4 \bar{\chi} + p^2 }{ 4 p^2} \right)^{1/2} + \frac{1}{2}
\right\}^2 \right] - 1 \right] \right\}
\nonumber
\end{eqnarray}
The divergence associated with
$B(p^2,\bar{\chi})$ may be absorbed in the renormalization of
$\lambda$ \cite{Root}. In fact, the expression
\begin{equation}
F( \bar{\chi}, p^2 ) \equiv
\left( p^2 + \bar{\chi} \right) \left( 3/\lambda_0 +
B_{T=0}( \bar{\chi}, p^2 ) \right) + \hat{\phi}^2  ,
\end{equation}
appearing in the potential, Eq.(\ref{eq:effpot}),
is proportional to the inverse
propagator of the massive scalar particle
$\sigma \equiv \phi_{1,1}$ and
may be rewritten as
\begin{equation}
F( \bar{\chi}, p^2 ) =
 \left( p^2 + \bar{\chi} \right) \left( 3/\lambda +
\bar{B}_{T=0}( \bar{\chi}, p^2 ) \right) + \hat{\phi}^2  ,
\end{equation}
where
\begin{eqnarray}
\bar{B}_{T=0}(p,\bar{\chi}(\hat{\phi}^2))
& = & \frac{1}{16 \pi^2}
\left\{ \ln \left( \frac{M^2}{\bar{\chi}}\right)
\right.
\nonumber\\
& - &
\left.
2 \left[ \left( \frac{ 4 \bar{\chi} + p^2 }{ 4 p^2} \right)^{1/2}
\ln \left[ \frac{p^2}{\bar{\chi}} \left\{ \left(
\frac{ 4 \bar{\chi} + p^2 }{ 4 p^2} \right)^{1/2} + \frac{1}{2}
\right\}^2 \right] - 1 \right] \right\}
\end{eqnarray}
and $\lambda$ is the renormalized quartic coupling at leading order
in $1/N$, which is given in
Eq.(\ref{eq:renlam}). Observe that, as first noticed in
Ref.\cite{CJP},
the renormalized theory seems to be spoiled by the presence of
tachyons. In fact, since we are working in Euclidean space,
a tachyon pole in the $\phi_{1,1}$ two point function will appear
if, for some positive
value of $p^2$, $F(\bar{\chi},p^2) = 0$. Considering the limit
 $\bar{\chi}
= \hat{\phi} = 0$, this implies an equation for $p^2$, which
reads
\begin{equation}
\frac{3}{\lambda} + \frac{1}{16\pi^2} \ln \left( \frac{ M^2}{p^2}
\right) + \frac{1}{8 \pi^2} = 0.
\label{eq:tach}
\end{equation}
For a finite value of the renormalized coupling $\lambda$, the
above equation has a solution at sufficiently large values of
$p^2$, which could be interpreted as the presence of a tachyon in
the spectrum. However, rewriting Eq.(\ref{eq:tach})  in
terms of the unrenormalized quartic coupling, we
obtain,
\begin{equation}
\frac{3}{\lambda_0} + \frac{1}{16 \pi^2} \ln \left(
\frac{ \Lambda^2 }{ p^2 } \right) + \frac{1}{8 \pi^2} = 0.
\label{eq:bare}
\end{equation}
Therefore, for any $\lambda_0 > 0$,
the potentially dangerous pole appears at momentum
above the cutoff scale \cite{BM},
\begin{equation}
p^2 \simeq \Lambda^2 \exp \left(48 \pi^2/\lambda_0
\right).
\end{equation}
Hence, it
has no physical consequences. (Observe
that at nonvanishing values of $\hat{\phi}$ and $\bar{\chi}$
the pole would be at even larger values of $p^2$.) The
apparent discrepancy between the results obtained in terms of
the renormalized and unrenormalized couplings is related to
the issue of triviality. In fact, while naively analyzing
the existence of a tachyon pole using Eq.(\ref{eq:tach}), one overlooks
the fact that the
renormalized
quartic coupling $\lambda$ goes to zero in the continuum limit
$\Lambda \rightarrow \infty$, and for this reason,
a finite effective cutoff
is needed in order to have a finite renormalized coupling.

An analogous situation occurs in the gauge sector of the theory.
In fact, calling
\begin{equation}
i D^{\mu \nu} = D_{{\cal{T}}}
\left( \delta^{\mu \nu} -
\frac{ p^{\mu} p^{\nu}}{p^2} \right) + D_L \frac{p^{\mu}
p^{\nu}}{p^2}
\label{eq:gendaa}
\end{equation}
and
\begin{equation}
\Pi^{\mu\nu}_{T=0}(\bar{\chi},p^2)
= \left( \delta^{\mu \nu} -
\frac{ p^{\mu} p^{\nu}}{p^2} \right)
 \Pi_{T=0} (\bar{\chi},p^2),
\label{eq:ppi}
\end{equation}
a tachyon pole in the current - current correlation function would
appear if
\begin{equation}
G\left( p^2,
\bar{\chi} \right) \equiv
\frac{1}{e_0^2} D_{{\cal{T}}}^{-1}
\left( \bar{\chi}, \hat{\phi}, p^2
\right) - \Pi_{T=0}
 \left( \bar{\chi},p^2 \right)
\end{equation}
had a  zero in Euclidean space. At  $\bar{\chi} =
\hat{\phi} = 0$, the condition $G(p^2, \bar{\chi}) = 0$
leads to the equation
\begin{equation}
\frac{1}{e_0^2} + \frac{1}{48 \pi^2} \left[ \ln \left(\frac{
\Lambda^2}{p^2}\right)
 + \frac{8}{3} \right] = 0 .
\end{equation}
Apart from a change from the gauge to the quartic
coupling, the above
 equation is  equivalent to that one
found in the previous case, Eq.(\ref{eq:bare}).
The would be tachyon pole appears
at momentum above the physical cutoff of the theory
\begin{equation}
p^2 \simeq \Lambda^2 \exp( 48 \pi^2/e_0^2 )
\end{equation}
and, hence, has no physical implications.

The renormalization of the gauge coupling proceeds in
similar way to that one of the
quartic coupling. The logarithmic divergences in
$\Pi^{\mu\nu}_{T=0}$ are absorbed in the definition of
the renormalized gauge coupling
$e$, which  is given by
\begin{equation}
\frac{1}{e^2} = \frac{1}{e_0^2} + \frac{ 1 }{ 48 \pi^2}
\ln \left( \frac{ \Lambda^2 }{ M^2 } \right).
\label{eq:rene}
\end{equation}
We can therefore define the renormalized vacuum polarization,
$\bar{\Pi}_{T=0}$, by replacing $\ln(\Lambda^2/\bar{\chi})$ by
$\ln(M^2/\bar{\chi})$ in $\Pi_{T=0}$. Moreover,
the expression of the renormalized gauge coupling,
Eq.(\ref{eq:rene}),
shows the triviality of the
gauged theory in the continuum
limit.

\subsection{ Radiative Corrections at Finite Temperature}

We shall now analyze the structure of $B(p,\bar{\chi})$ and
$\Pi^{\mu \nu}(p, \bar{\chi})$ at finite temperature,
which, by making use of Eq.(\ref{eq:FTtrick}), may
be decomposed into zero temperature contributions,
$\bar{B}_{T=0}$ and $\Pi^{\mu\nu}_{T=0}$, already
analyzed in section 5.1,
and the corresponding
finite temperature parts  to be computed in the following.
The radiative correction to the renormalized $\chi$
propagator, $\bar{B}_T(p,\bar{\chi})$, has the expression
\begin{eqnarray}
\bar{B}_T(p,\bar{\chi}) - \bar{B}_{T=0}(p,\bar{\chi}) & = &
\int_{-\infty + i \epsilon}^{\infty + i \epsilon}
\frac{dk_0}{ 2 \pi }
\int \frac{ d^3 k}{ ( 2 \pi )^3 }
\left\{ \frac{ 1 }
{ \left[ k^2 + \bar{\chi} \right] \left[ (k + p)^2
+ \bar{\chi} \right] } \right.
\nonumber\\
& + &
\left.
\frac{ 1 }
{ \left[ \tilde{k}^2 + \bar{\chi} \right] \left[ (\tilde{k} + p)^2
+ \bar{\chi} \right] }
\right \} \frac{1}{ \left( \exp(-i \beta k_0) - 1 \right)}
\end{eqnarray}
where $\tilde{k} = ( -k_0, \vec{k})$. The $k_0$ integral may
be performed by making a contour integration in the complex
$k_0$ plane, leading to
\begin{eqnarray}
\bar{B}_T(p, \bar{\chi}) & = & \bar{B}_{T=0}(p,\bar{\chi}) +
2 \int \frac{d^3 k}{ (2 \pi)^3 } \left[ \frac{1}{
\sqrt{\vec{k}^2 + \bar{\chi} }}
\frac{1}{ \left[ \exp\left( \beta \sqrt{ \vec{k}^2 + \bar{\chi} } \right)
- 1 \right]}
\right.
\nonumber\\
& &  \left.
\times
\frac{ \left( \vec{k} + \vec{p} \right)^2 - \vec{k}^2
+ p_0^2 }{ \left\{  \left[ \left( \vec{k} + \vec{p} \right)^2 -
\vec{k}^2 + p_0^2 \right]^2 + 4 p_0^2 \left( \vec{k}^2 + \bar{\chi}
\right) \right\} } \right].
\label{eq:BT}
\end{eqnarray}

As we have shown above,
in order to study the
minimization condition up to next to leading order in $1/N$,
it is sufficient  to study the
behaviour of $\bar{B}_T$ for $\bar{\chi} \simeq 0$.
{}From the expression
of $\bar{B}_T(p,\bar{\chi})$,
doing an expansion in the neighborhood of $\bar{\chi} =0$
at  finite
external momentum, we obtain
\begin{equation}
\bar{B}_T (p_0, \vec{p}, \bar{\chi}) =
\bar{B}_T(p_0, \vec{p}, \bar{\chi} = 0 ) +
K \frac{ \sqrt{ \bar{\chi} }}{ \beta p^2} + {\cal{O}}( \bar{\chi}),
\label{eq:above}
\end{equation}
where $K$ is a coefficient of order one, which can be evaluated by
analyzing the infrared divergences associated with $\partial
\bar{B}_T/\partial \bar{\chi}$ as $\bar{\chi} \rightarrow 0$.
{}From Eq.(\ref{eq:BT})
it is possible to prove that
\begin{equation}
\frac{ \partial \bar{B}_T(p,\bar{\chi})
}{ \partial \sqrt{\bar{\chi}} } \rightarrow
- \frac{1}{2 \pi \beta \left( p_0^2 + \vec{p}^2 \right) }
\end{equation}
as $\bar{\chi} \rightarrow 0$ and, hence,
$K = - 1/2\pi$.
Furthermore, the last term in the
Eq.(\ref{eq:above}) involves higher order contributions
in $\bar{\chi}$ which
are not relevant to solve the minimization condition.
This is due to
the fact that, from the gap equation up to
leading order in 1/N, Eq.(\ref{eq:leadgap2}),
\begin{equation}
\left.
\frac{\partial \bar{\chi}(\hat{\phi}^2)}{\partial
\hat{\phi}^2} = \frac{ 4 \pi \beta \sqrt{\bar{\chi}}}
{1 + 24 \pi \beta \sqrt{\bar{\chi}}/\lambda_T}
\right|_{\bar{\chi}
\rightarrow 0}
\rightarrow
4 \pi \beta \sqrt{\bar{\chi}}.
\end{equation}
Therefore, the  derivative
operator  involved in the
minimization condition, Eq.(\ref{eq:min}),
only receives contributions from the infrared dominant
part of
$\partial V_{n.l.o} / \partial \bar{\chi} $.

An analogous procedure may be used to analyze the behaviour
of the
vacuum polarization contribution
$\Pi^{\mu\nu}_T$.
\begin{eqnarray}
\Pi^{\mu \nu}_T - \Pi^{\mu \nu}_{T=0}  & = &
\int_{-\infty + i \epsilon}^{\infty + i \epsilon}
 \frac{dk_0}{2\pi}
\int \frac{ d^3 k}{ ( 2 \pi )^3 }
\left\{ \frac{ ( 2 k + p )^{\mu} ( 2 k + p )^{\nu} }
{ \left[ k^2 + \bar{\chi} \right] \left[ (k + p)^2
+ \bar{\chi} \right] } \right.
\nonumber\\
& + &
\left.
\frac{ ( 2 \tilde{k} + p )^{\mu} ( 2 \tilde{k} + p )^{\nu} }
{ \left[ \tilde{k}^2 + \bar{\chi} \right] \left[ (\tilde{k} + p)^2
+ \bar{\chi} \right] } -
4 \frac{ \delta^{\mu \nu} }{ k^2 + \bar{\chi} }
\right \} \frac{1}{ \left( \exp(-i \beta k_0) - 1 \right)}
\label{eq:pimunut}
\end{eqnarray}
At finite temperature it is easy to
demonstrate the transversality of $\Pi^{\mu \nu}_T$ by making
use of the $p_0$ quantization,
$p_0 \equiv \omega_n = 2 \pi n
/ \beta_0$. However, general covariance is explicitly lost and,
hence, the vacuum polarization takes the general form\cite{GPY}
\begin{equation}
\Pi^{\mu \nu}_T(p, \bar{\chi}) = \Pi_1^T(p, \bar{\chi})
\left( \delta^{\mu \nu} -
\frac{p^{\mu} p^{\nu}}{p^2} \right) + \Pi_2^T(p, \bar{\chi})
\left( \delta^{i j} - \frac{ p^i p^j}{\vec{p}^2} \right).
\label{eq:cov}
\end{equation}
Recalling the general form of the gauge boson propagator,
Eq.(\ref{eq:gendaa}), it is straightforward to show that
\begin{eqnarray}
\det
\left( -i D_{\mu \nu}^{-1} - e_0^2 \Pi_{\mu \nu}^T \right) & = &
\left[ \frac{1}{D_{{\cal{T}}}
(p^2,\hat{\phi}) } - e_0^2 \left( \Pi_1^T(p,
\bar{\chi})
+ \Pi_2^T(p,\bar{\chi})
\right) \right]^2
\nonumber\\
& & \left[ \frac{1}{
D_{{\cal{T}}}(p^2,\hat{\phi})}
- e_0^2 \Pi_1^T (p,\bar{\chi})
\right] \frac{1}{D_L(p^2,\bar{\chi},\hat{\phi})} ,
\label{eq:det}
\end{eqnarray}
where, from  Eq.(\ref{eq:daa}), we
have,
\begin{eqnarray}
\frac{1}{D_{{\cal{T}}}
(k^2,\hat{\phi})} & = & k^2 + e_0^2 \hat{\phi}^2
\nonumber\\
\frac{1}{D_L(k^2,\bar{\chi},\hat{\phi})} & = &
\frac{\left[k^2 \left( k^2 + \bar{\chi} \right) +
\alpha e_0^2 \hat{\phi}^2 \bar{\chi} \right]}
{ \alpha \left( k^2 + \bar{\chi} \right) }.
\end{eqnarray}
Therefore,
\begin{eqnarray}
\det \left( -i D_{\mu\nu}^{-1} -  e_0^2 \Pi_{\mu\nu}^T
\right) & = &  \left[ k^2 + e_0^2 \hat{\phi}^2 -
e_0^2 \left( \Pi_1^T(k,\bar{\chi}) + \Pi_2^T(k,\bar{\chi})
\right) \right]^2
\nonumber\\
& &
\left[k^2 + e_0^2 \hat{\phi}^2 -
e_0^2 \Pi_1^T(k, \bar{\chi}) \right] \frac{1}{D_L} .
\label{eq:determi}
\end{eqnarray}

This expression is equivalent to that one found by
Carrington \cite{improved3} in the limit of
small external momentum
$(p_0 = 0,\vec{p} \simeq 0)$, while computing the ring diagrams
contribution to the effective potential in the Landau gauge.
Observe that, at zero temperature, $\Pi_1^{T=0} =
\Pi_{T=0}$,  while
$\Pi_2^{T=0} = 0$ and all logarithmic divergences may
be absorbed in the definition of the renormalized gauge
coupling, which implies that
$\Pi_1^{T=0} = \Pi_{T=0}$, turns into its
renormalized expression
$\bar{\Pi}_1^{T=0} = \bar{\Pi}_{T=0}$. Consequently,
defining $\bar{\Pi}_1^T$ as the finite temperature
renormalized
quantity,
the logarithm of the determinant takes
the form
\begin{eqnarray}
\ln \det [-i D^{-1}_{\mu \nu}(k,\hat{\phi},\bar{\chi})
&-&
e_0^2 \Pi_{\mu \nu}^T ( k, \bar{\chi})]  =
2 \ln \left[ k^2 + e^2 \hat{\phi}^2 - e^2 \left( \bar{\Pi}_1^T
+ \Pi_2^T \right) \right]
\nonumber\\
& + & \ln \left[ k^2 + e^2 \hat{\phi}^2
- e^2 \bar{\Pi}_1^T \right] + \ln \left\{
\frac{k^2 \left( k^2 + \bar{\chi} \right) +
\alpha e_0^2 \hat{\phi}^2 \bar{\chi}}
{  k^2 + \bar{\chi}  } \right\}.
\label{eq:lndet}
\end{eqnarray}
The first two terms in the right hand side of the equation
above
are gauge independent and, for $\Pi_i^T = 0$, they
give the well known one loop contribution to the effective potential.
The last term
contains all the gauge dependence and its
contribution to the effective potential only vanishes in the
Landau gauge, $\alpha = 0$. We shall return to the
issue of the gauge dependence in section 5.3.

Since $\Pi_{\mu \nu}^T$, Eq.(\ref{eq:cov}),
has only two independent components,
we can obtain the functions $\bar{\Pi}_i^T$
by computing the expressions
of $\Pi_{00}^T$ and $\Pi_{\mu\mu}^T$. In fact,
\begin{eqnarray}
\Pi_{00}^T - \Pi_{00}^{T=0} & = &
\left( \bar{\Pi}_1^T - \bar{\Pi}_1^{T=0} \right)
\frac{ \vec{p}^2 }{ p^2 } ,
\nonumber\\
\Pi^{\mu \mu}_T - \Pi^{\mu \mu}_{T=0} & = &
3 \left( \bar{\Pi}_1^T - \bar{\Pi}_1^{T=0}
\right) + 2  \Pi_2^T .
\end{eqnarray}
Moreover, using
Eqs.(\ref{eq:pimunut0}) and (\ref{eq:pimunut}), we have
\begin{eqnarray}
\Pi^{00}_T(k,\bar{\chi}) - \Pi^{00}_{T=0}(k,\bar{\chi})
& = & I_1(k,\bar{\chi}) - 2 I_2(k,\bar{\chi})
\label{eq:ionetwo}
\\
\Pi^{\mu \mu}_T(k,\bar{\chi})
 - \Pi^{\mu \mu}_{T=0}(k,\bar{\chi})
& = &
-4 I_2(k,\bar{\chi}) - \left( k^2 + 4 \bar{\chi} \right)
\left( \bar{B}_T(k,\bar{\chi},\hat{\phi}) -
\bar{B}_{T=0}(k,\bar{\chi},\hat{\phi}) \right) ,
\nonumber
\end{eqnarray}
with
\begin{eqnarray}
I_1(p,\bar{\chi}) & = &
\int_{-\infty + i \epsilon}^{\infty +
i \epsilon} \frac{dk_0}{2\pi} \int
\frac{d^3k}{(2 \pi)^3}
\left\{
\frac{ \left(2 k_0 + p_0 \right)^2 }{ \left( k^2 +
\bar{\chi} \right) \left[ \left( k + p \right)^2 +
\bar{\chi} \right]} \right.
\nonumber\\
& + &  \left.
\frac{ \left(2 k_0 - p_0 \right)^2 }{ \left( \tilde{k}^2 +
\bar{\chi} \right) \left[ \left( \tilde{k} + p \right)^2 +
\bar{\chi} \right]}\right\} \frac{1}{ \left( \exp(-i \beta k_0)
-1 \right) }
\end{eqnarray}
and
\begin{equation}
I_2(p,\bar{\chi})
= 2
\int_{-\infty + i \epsilon}^{\infty +
i \epsilon} \frac{dk_0}{2\pi} \int
\frac{d^3 k}{(2 \pi)^3}
\frac{1}{
\left( k^2 + \bar{\chi} \right)} \frac{1}{ \left( \exp(-i \beta k_0)
-1 \right) }.
\end{equation}

The integrals $I_i$ may be computed by performing a contour
integration in the complex $k_0$ plane. We obtain,
\begin{eqnarray}
I_1 & = & 2 \int \frac{d^3 k}{( 2 \pi)^3} \frac{1}{
\sqrt{ \vec{k}^2 + \bar{\chi}} \left[ \exp \left(\beta
\sqrt{ \vec{k}^2 + \bar{\chi}} \right) - 1 \right]}
\nonumber\\
&&
\times
\frac{\left[ p_0^2
- 4 \left( \vec{k}^2 + \bar{\chi} \right) \right]
\left[ p_0^2 + \left( \vec{k} + \vec{p} \right)^2
- \vec{k}^2 \right] + 8 p_0^2 \left( \vec{k}^2
+ \bar{\chi} \right)}{ \left[ p_0^2 + \left( \vec{k}
+ \vec{p} \right)^2 - \vec{k}^2 \right]^2
+ 4 p_0^2 \left( \vec{k}^2 + \bar{\chi} \right) }
\label{eq:ione}
\end{eqnarray}
and
\begin{equation}
I_2 = \int \frac{d^3 k}{(2 \pi)^3}
\frac{1}{ \sqrt{ \vec{k}^2 + \bar{\chi} }
\left( \exp \left( \beta \sqrt{ \vec{k}^2 + \bar{\chi} }
\right)  - 1
\right) } .
\label{eq:itwo}
\end{equation}

It is first interesting to analyze the behaviour of
$I_i (p_0, \vec{p} )$ at $\bar{\chi} = 0$, $p_0 = 0$ and
$|\vec{p}| \ll T$. The integral $I_2$ has already been
considered in section 4, and its high temperature expansion
result, Eq.(\ref{eq:DJ}), in the limit we are studying, gives
$I_2(\bar{\chi}=0) = 1/12\beta^2$. Furthermore, from
Eq.(\ref{eq:ione}),
it is straightforward to show that
\begin{equation}
I_1( \bar{\chi} = 0, p_0 = 0, \vec{p} \simeq 0 )
= - \frac{1}{\pi^2} \int_0^{\infty} d|\vec{k}|
\frac{|\vec{k}|}{
\left( \exp( \beta |\vec{k}|) - 1 \right)}
= - \frac{1}{ 6 \beta^2}.
\end{equation}
The above results  imply that, in this momentum regime,
\begin{equation}
 \bar{\Pi}_1^T - \bar{\Pi}_1^{T=0} =
- \Pi_2^T =
- \frac{1}{3 \beta^2}
\end{equation}
Observe that this contribution gives a static mass term to the
abelian gauge field. In fact, recalling
Eq.(\ref{eq:determi}) we obtain that, since
$\bar{\Pi}_1^T$ and $\Pi_2^T$
come with opposite signs,
only the longitudinal
mode receives a contribution to the mass term equal to
\begin{equation}
{\cal{M}}_D^2 = \frac{e^2}{3 \beta^2},
\end{equation}
which is the well known Debye screening mass term \cite{GPY}.
It has been
recently pointed out in the literature
\cite{improved3}-\cite{improved5},
that
this term is important in order to
determine the strength of the phase transition
within the framework of the
improved one loop effective potential.
Within such approximation,
the phase transition appears
to be first order but, due to the Debye screening suppression of
the longitudinal mode, its strength is about two thirds of the
one obtained using the naive one loop approach. We shall now
demonstrate that, when using the self consistent 1/N expansion up
to next to leading order in 1/N, and computing the value of
$\Pi^{\mu\nu}_T$ at finite external momentum,
extra  terms
appear in the effective potential,
which change not only the
strength but also the order of the phase transition.

\subsection{Minimization of the Effective Potential}

The full effective potential up to next to leading order in
1/N, Eq.(\ref{eq:effpot}) may be rewritten as
\begin{eqnarray}
V( \bar{\chi}, \hat{\phi} ) &  = &V_{tree} +
V_{l.o.} + \frac{1}{2} \int_k \ln \left[
\frac{ \left( k^2 + \bar{\chi} \right)
\left( 1 + \lambda \bar{B}_T (\bar{\chi},k )/3
\right) + \lambda \hat{\phi}^2/3 } { k^2 + \bar{\chi} }
\right]
\nonumber\\
& + & \frac{1}{2} \int_k \left\{
2 \ln \left[ k^2 + e^2 \hat{\phi}^2 - e^2 \left(
\bar{\Pi}_1^T
+ \Pi_2^T \right) \right] \right.
\nonumber\\
& + & \left.
\ln \left[ k^2 + e^2 \hat{\phi}^2
- e^2 \bar{\Pi}_1^T \right]  \right\}  + V_{g.d.} ,
\label{eq:fulleff}
\end{eqnarray}
where $e$ and $\lambda$ are the renormalized gauge and
quartic couplings, respectively,
 and $\bar{B}^T$ and $\bar{\Pi}_i^T$ are the
full, temperature dependent radiative corrections contributions
with
their logarithmic divergences
removed through the definition of
the renormalized couplings, in the way explained above.
Moreover, we can rewrite the gauge dependent term, $V_{g.d.}$,
coming from Eq.(\ref{eq:lndet}) like
\begin{equation}
V_{g.d.} = \frac{1}{2} \int_k \left[
\ln \left( k^2 + R_1^2 \right)  + \ln \left( k^2 + R_2^2 \right)
- \ln \left( k^2 + \bar{\chi} \right) \right]
\end{equation}
where
\begin{equation}
R_{1,2}^2 = \frac{ \bar{\chi} \pm \sqrt{ \bar{\chi}^2 -
4 \alpha e_0^2 \hat{\phi}^2 \bar{\chi} } }{2}.
\end{equation}

Similarly to  the computation of $\bar{B}_T$, from
Eq.(\ref{eq:min}) it follows that, in order to evaluate
the minimum condition up to next to leading order in
$1/N$, we only need to study the behaviour of
$\bar{\Pi}_i^T$
for $\bar{\chi} \simeq 0$.
As we have
just remarked, the computation of the vacuum polarization
expression for finite external momentum is crucial in
obtaining the correct behaviour of the effective potential
at small values of $\bar{\chi}$.
Recalling Eqs.(\ref{eq:pimunut0}),(\ref{eq:pimunut}),
(\ref{eq:ionetwo}),(\ref{eq:ione}) and (\ref{eq:itwo}),
it is straightforward to
find that, for $\bar{\chi} \rightarrow 0$, $p^{\mu} \neq 0$,
\begin{equation}
\Pi^{\mu\mu}_T(p_0,\vec{p},\bar{\chi})  =
\Pi^{\mu\mu}_T(p_0,\vec{p},\bar{\chi} = 0) +
\frac{ 3 \sqrt{ \bar{\chi}} }{ 2 \pi \beta} + h.o.(\bar{\chi}),
\label{eq:ppmm}
\end{equation}
and
\begin{equation}
\Pi^{00}_T(p_0,\vec{p},\bar{\chi}) =
\Pi^{00}_T(p_0,\vec{p},\bar{\chi} = 0) +
\frac{\sqrt{\bar{\chi}}}{2 \pi \beta} \frac{\vec{p}^2}
{\left(p_0^2 +
\vec{p}^2 \right)} + h.o.(\bar{\chi}).
\end{equation}
Note that, as implied by transversality,
the expression of $\Pi_{00}^T$
 vanishes in the limit $\vec{p} \rightarrow 0$
for $p_0 \neq 0$.
{}From the above expressions it follows that, at finite
momentum transfer and in the small $\bar{\chi}$ limit,
the vacuum polarization components read,
\begin{eqnarray}
\bar{\Pi}_1^T(p_0,\vec{p},\bar{\chi}) & = &
\bar{\Pi}_1^T(p_0,\vec{p},\bar{\chi} = 0)
+ \frac{\sqrt{\chi}}{2 \pi \beta} + h.o.(\bar{\chi})
\nonumber\\
\Pi_2^T(p_0,\vec{p},\bar{\chi}) & = &
\Pi_2^T(p_0,\vec{p},\bar{\chi} = 0)
+ h.o.(\bar{\chi}).
\label{eq:asympi}
\end{eqnarray}
It is important to emphasize that, this is the correct behaviour
for $\bar{\chi} \rightarrow 0$
at any finite $|p| = \sqrt{p_0^2 + \vec{p}^2}$, which is the
relevant regime for the computation of the integrals contributing
to the minimum of the effective potential.
Using the above expressions, Eqs.(\ref{eq:above}),(\ref{eq:asympi}),
we can rewrite the minimization condition as
\begin{eqnarray}
\bar{\chi} & = & - \frac{1}{N} \left[ \frac{\partial}{\partial
\hat{\phi}^2} + \frac{\partial \bar{\chi}}{\partial
\hat{\phi}^2} \frac{\partial}{\partial \bar{\chi}} \right]
\left\{ V_{g.d.}
\right.
\nonumber\\
& + &
\int_k \ln \left[ k^2 + e^2 \hat{\phi}^2 -
e^2 \left( \bar{\Pi}_1^T(k,\bar{\chi}=0) + \frac{
\sqrt{\bar{\chi}}}
{2 \pi \beta} \right) \right]
\nonumber\\
& + &
2 \int_k \ln \left[ k^2 + e^2 \hat{\phi}^2 -
e^2 \left( \bar{\Pi}_1^T(k,\bar{\chi}=0) +
\Pi_2^T(k,\bar{\chi}=0)  +
\frac{\sqrt{\bar{\chi}}}
{2 \pi \beta} \right) \right]
\nonumber\\
&+& \left.
\int_k \ln \left[ \frac{ \left( k^2 + \bar{\chi} \right)
\left( 3/\lambda + \bar{B}_T(k, \bar{\chi}=0) -
\sqrt{\bar{\chi}}/(2\pi\beta k^2) \right) + \hat{\phi}^2}{k^2 +
\bar{\chi}} \right] \right\}_{\bar{\chi} = 0}
\label{eq:chimin}
\end{eqnarray}
The first term in Eq.(\ref{eq:chimin}) involves all the possible
gauge dependent contributions to
 the minimum of the effective potential.
However, explicitly applying the total derivative to the
logarithmic  expression  of $V_{g.d.}$,
\begin{eqnarray}
\left.
\left( \frac{\partial}{\partial \hat{\phi}^2 } +
\frac{ \partial \bar{\chi} }{ \partial \hat{\phi}^2 }
\frac{ \partial }{ \partial \bar{\chi} }\right) V_{g.d.}
\right|_{\bar{\chi}=0} & = &
\frac{1}{2} \left\{ \int_k \frac{1}{k^2} \left[
\frac{\partial}{\partial \hat{\phi}^2 } \right. \right.
\nonumber\\
& + &   \left. \left.
\frac{ \partial \bar{\chi} }{ \partial \hat{\phi}^2 }
\frac{ \partial }{ \partial \bar{\chi}} \right]
\left( R_1^2 + R_2^2 - \bar{\chi} \right) \right\}_{\bar{\chi} = 0}
\end{eqnarray}
and
using the fact that $\left( R_1^2 + R_2^2 - \bar{\chi} \right) = 0$,
we observe that $V_{g.d.}$ does not give any contribution to the
minimization condition up to this order in $1/N$.
( Observe that, since
we are working with regularized integrals with a finite cutoff,
we can safely interchange derivatives by integrals in the expression
above.) Therefore, although
the expression of the
effective potential is gauge dependent, the minimization
condition gives a gauge independent result.

Furthermore, from Eq.(\ref{eq:chimin}), it comes to notice
that, as we discussed above, for the minimization of the
effective potential,
the only relevant contributions from the finite
temperature radiative corrections
come from the infrared dominant terms
of  $\bar{B}_T(p,\bar{\chi})$ and
$\bar{\Pi}_i^T(p,\bar{\chi})$. In addition,
recalling the fact that $\partial \bar{\chi}/ \partial
\hat{\phi}^2 \rightarrow  \sqrt{\bar{\chi}} 4 \pi \beta$ as
$\bar{\chi} \rightarrow 0$,
it readily follows that the minimization condition reads,
\begin{eqnarray}
\bar{\chi} & = & - \frac{1}{N} \left\{ 2
e^2 \int_k
\frac{ 1 - 2 \sqrt{\bar{\chi}} \partial(\sqrt{\bar{\chi}})/
\partial \bar{\chi} }{\left[
k^2 + e^2 \hat{\phi}^2 - e^2
\left( \bar{\Pi}_1^T(k,\bar{\chi}=0) + \Pi_2^T(k,\bar{\chi}=0)
\right)\right]}  \right.
\nonumber\\
& + & e^2 \int_k
\frac{ 1 - 2 \sqrt{\bar{\chi}} \partial(\sqrt{\bar{\chi}})/
\partial \bar{\chi} }{
\left[ k^2 + e^2 \hat{\phi}^2 - e^2
\left( \bar{\Pi}_1^T(k,\bar{\chi}=0)\right)
\right]}
\nonumber\\
& + &  \left.
\int_k \frac{ 1 - 2 \sqrt{\bar{\chi}} \partial (\sqrt{\bar{\chi}})/
\partial \bar{\chi} }{ \left[ k^2 \left( 3/\lambda +
\bar{B}_T(k,\bar{\chi}=0) \right) + \hat{\phi}^2 \right] }
\right\}_{\bar{\chi}=0}
\end{eqnarray}
The above equation implies that,
at next to leading order in $1/N$ the only
minimum is at
\begin{equation}
\bar{\chi}_{min}(\hat{\phi}) = 0.
\label{eq:vanchi}
\end{equation}
Thus, the value of $\bar{\chi}(\hat{\phi})$
at the minimum remains unchanged
after the inclusion of the next to leading order contributions.
As we discuss in section 4,
Eq.(\ref{eq:vanchi}) leads to a
 nontrivial solution for
the vacuum expectation value of the scalar field,
$\hat{\phi}_{min}^2 =
v^2 - 1/(6 \beta^2)$,
for temperatures
below the critical temperature, $T_c^2 = 6 v^2$.
Hence, the vacuum expectation value of the
scalar field varies continuously at the critical temperature,
characterizing a second order phase transition.

\section{Comparison with the One Loop Approximation}

 The results of  section 5 show that, in the presence
of a large number of scalars and in the region of parameters
$e^2/\lambda \ll N$,
in which the scalar loop contributions
are enhanced in comparison to the gauge loop ones, the
phase transition is second order.
In spite of that,  the gauge contributions  to the
effective potential at next
to leading order in $1/N$ resemble those ones contributing
to the effective potential in
the improved one loop approximation.
In fact, for small
$\hat{\phi}$, and at $T = T_0$, at which the quadratic
term in $\hat{\phi}$ vanishes, one would expect that due
to  the presence
of the  gauge contributions, a dominant
negative cubic term  would appear,
 leading to a nontrivial
minimum in the effective potential and, hence, to a first order
phase transition.  However,
the fundamental difference
comes from considering the finite external momentum contributions
to the radiative corrections at finite temperature.

Quite generally, as we have already said, the integral
\begin{eqnarray}
I_T  =  \frac{1}{2} \int_k \ln(k^2 + m^2)
\label{eq:last}
\end{eqnarray}
may be evaluated in the high temperature regime to give,
\begin{eqnarray}
I_T  =  I_{T=0} + \frac{m^2}{24 \beta^2} - \frac{m^{3/2}}{12 \pi
\beta} + h.o.(m^2\beta^2).
\label{eq:intt}
\end{eqnarray}
Therefore,
since the zero temperature gauge boson mass is equal to
$m_g^2 = e^2 \hat{\phi}^2$ one would naively
expect  that the inclusion of the gauge field would induce
the generation
of a quadratic term in $\hat{\phi}$, which would modify the
definition of the
transition temperature, as well as the appearance of a
 cubic term, which would lead to
a first order phase transition. In the next to leading
order in $1/N$ analysis,
these two effects are not present, and, hence,
some sort of finite temperature screening,
which modifies the nature of the phase
 transition, should occur.
One type   of screening, which is already contained
within the improved one loop approximation, is that one
induced by the generation of an effective temperature
dependent mass for the longitudinal gauge boson mode. As
we already discussed above, when compared with the naive
one loop result,
the Debye screening causes the
reduction in two thirds of the value of the coefficient
of the cubic term. In our analysis,
however, the preservation of the value of the critical
temperature and of the nature of the phase transition
after considering the
next to leading order in 1/N
effects is a result that
would be obtained even in the case of a cancellation
of the $\bar{\chi}(\hat{\phi})$-independent  radiative
corrections contributions to the photon and scalar propagators.
In fact, within the 1/N expansion, there is a different source
of screening, which can only be obtained by considering the
$\hat{\phi}$ dependent, finite momentum contributions to the
radiative corrections.

For values of $\hat{\phi}$ close to the minimum of the effective
potential ($\bar{\chi} \simeq 0$),
and at momentum $p^2 \gg \bar{\chi}$, the evaluation of the
finite temperature
radiative corrections to the photon and scalar propagators
have as outcome the replacement of the functional
dependence on $\hat{\phi}^2$ by a dependence on the combination
$[\hat{\phi}^2 - \sqrt{\bar{\chi}}/( 2 \pi \beta)]$.
 In fact, as it
is clearly seen from Eqs.(\ref{eq:fulleff}),
(\ref{eq:ppmm})-(\ref{eq:asympi}),
in this momentum regime, the
finite temperature effective mass of the transverse gauge
boson mode is given by
\begin{equation}
m^2_{{\cal{T}}}
= e^2 \left({\phi}^2 - \sqrt{\bar{\chi}}/(2 \pi \beta)\right)
+ h.o.(\bar{\chi})
\end{equation}
To easily understand
the relevance of this effect, let us first
analyze
the behaviour of $\bar{\chi}$ at $T = T_c$,
\begin{equation}
\sqrt{\bar{\chi}} = - \frac{\lambda_{T_c}}{ 24 \pi \beta_c} +
\sqrt{ \left( \frac{\lambda_{T_c}}{ 24 \pi \beta_c} \right)^2 +
\frac{ \lambda_{T_c} \hat{\phi}^2 }{6} }.
\label{eq:chitc}
\end{equation}
Expanding Eq.(\ref{eq:chitc}) in the neighborhood of the origin
we obtain
\begin{equation}
\sqrt{\bar{\chi}} = 2 \pi \beta_c \hat{\phi}^2 \left[
1 -
\frac{24 \pi^2}{\lambda_{T_c}}\left(\frac{\phi}{T_c}\right)^2
\right].
\label{eq:propdep}
\end{equation}
{}From
Eq.(\ref{eq:propdep}) it follows that at $T = T_c$ and for
the momentum regime under consideration, the effective
mass of the photon transverse mode in the
neighborhood of the origin
is given by
\begin{equation}
m^2_{{\cal{T}}} =  \hat{\phi}^2
\frac{24 \pi^2 e^2}{\lambda_{T_c}} \left(
\frac{\hat{\phi}}{T_c}\right)^2 + h.o.(\bar{\chi})
\label{eq:cancel}
\end{equation}
Thus, since for the same range of parameters,
\begin{equation}
\bar{\chi} \simeq 4 \pi^2 \hat{\phi}^2 \left(
\frac{\hat{\phi}}{T_c}
\right)^2,
\label{eq:same}
\end{equation}
the effective squared
mass of the photon transverse mode behaves like
$\hat{\phi}^4/T^2$
rather than the expected $\hat{\phi}^2$ dependence.
Within the region of parameters considered above,
the radiatively corrected
squared mass of the $\sigma$ particle also behaves as
$\hat{\phi}^4/T^2$.

The dependence of $m_{{\cal{T}}}^2$ in the neighborhood of the origin
and at $T = T_c$ has two effects. First, it explains the
cancellation of the quadratic term in $\hat{\phi}$
at $T = T_c$ and hence
the preservation of the value of the
critical temperature at next to
leading order in $1/N$. Second, it
implies that, in the neighborhood of the origin,
and  independently
of the existence of
$\hat{\phi}$ - independent radiative
corrections contributions,
no cubic term is induced and thus,
at $T = T_c$ the finite
temperature effective potential is well described by a
positive, quartic term.

It is important to remark that the expressions
we obtain for the
radiatively corrected photon and scalar propagators are
only valid for $T, k^2 \gg \bar{\chi}$. Moreover, the cubic
term arising in the
integral, Eq.(\ref{eq:last}), receives the most relevant
contributions from the momentum regime $k_0 = 0$,
$\vec{k}^2 = {\cal{O}}(m^2)$.
Thus, the present analysis
would fail if in the neighborhood of the origin
the relation $\hat{\phi}^2 \gg \bar{\chi}$ were not
fulfilled.  However, as readily seen from Eq.(\ref{eq:same}),
$\bar{\chi} \ll \phi^2$ and  the effective
potential close to the origin is well described
by our approach.

\section{Conclusions}

In this article we have analyzed the finite temperature
phase transition of scalar electrodynamics with N
flavors of complex scalar fields, by means of a large N
expansion, up to next to leading order in $1/N$.
 We have shown that
the effective potential takes a compact and simple
form, which, although gauge dependent, leads to
gauge invariant results for the critical temperature
as well as for the extrema of the potential, which are
well defined physical quantities.
At leading order in $1/N$,
the effective potential coincides
with the one of the O(2N) vector model, already studied
in the literature. At the critical temperature $T_c = 6 v^2$,
the system develops a symmetry restoration
phase transition, with an order parameter which varies
continuously from zero to nonzero values. The phase  transition is
hence  second order.

At next to leading order in the $1/N$ expansion, the effective
potential receives contributions from
the radiatively corrected $\sigma$
and  gauge field  propagators.
The gauge field contribution is essentially that
one found at one loop, but with the photon corrected by
vacuum polarization effects. The $\hat{\phi}$ dependence of
the vacuum polarization effects is crucial  in order to
define a correct $1/N$ expansion. We have shown that, when
the radiative corrections are considered at finite external
momentum, their effects lead to an effective screening of
the gauge boson mass, which change the nature of the
gauge boson loop contributions to the effective potential.
Due to this screening effect,
no extra $\hat{\phi}^2$ term is induced in the neighborhood
of the origin at the leading order critical temperature
and, therefore, the critical temperature remains the same as
in the leading order analysis.
More generally, the position  of the minima of the effective
potential is not modified by next to leading order effects and,
in addition,
no new minimum of the effective potential appears
at next to leading order. Hence, the order parameter has
the same behaviour as in leading order in $1/N$ and the
phase transition remains second order.

The results of this work show that in the presence of a
large number of
flavors of complex scalar fields,
the symmetry restoration phase transition
of scalar QED becomes second
order. Within the context of this study, however, we
can not rule out the
presence of a critical value of $N$, below which the transition
is  first order.  In general, we expect the range of
validity of the large $N$ approximation to depend on the relation
between the gauge and quartic couplings.  More specifically,
the expansion considered in the present work
assumes the dominance
of the scalar loops contributions to the effective potential.
Therefore, it is
rigorously valid for values of $e^2/\lambda \ll N$.
Thus, for low values of $N$, the expansion becomes more
reliable for a scalar Higgs mass,
$m_{\sigma}^2 = \lambda \hat{\phi}^2/3$
 larger than the gauge boson
mass, $m_g^2 = e^2 \hat{\phi}^2$,
$m_{\sigma}^2/m_g^2 \geq 1$, for which the fluctuations of the scalar
fields are enhanced.
{}~\\
{}~\\
{}~\\
ACKNOWLEDGEMENTS

We would like to thank  W. A. Bardeen, R. D. Peccei,
R. D. Pisarski and P. Weisz
for very enjoyable discussions and helpful
comments.
We would also like to thank V. Jain for pleasant and
useful discussions
and P. Ramond and S. Hsu for interesting conversations.
Parts of this work were carried out at the Aspen Center for
Physics, at Fermilab National Laboratory and at Brookhaven
National Laboratory, to which we are grateful.
\newpage
FIGURE CAPTIONS\\
{}~\\
{}~\\
Fig.1. Multiloop diagrams which contribute to the effective
potential at next to leading order in $1/N$. Solid and dashed
lines denote the propagators of the scalar fields, $\phi_{a,i}$
and $\chi$, respectively, while the curved lines denote the
gauge field propagator.
{}~\\
{}~\\
Fig.2. Same as in Fig. 1, but considering those diagrams which
involve a mixing of $\chi$ and $A_{\mu}$.

\end{document}